\documentstyle[12pt]{article}

\begin{document}

\begin{flushright}
IMSc/2012/11/18 
\end{flushright} 

\vspace{2mm}

\vspace{2ex}

\begin{center}

{\large \bf Remarks on Black Hole Evolution } \\ 

\vspace{2ex}

{\large \bf a la Firewalls and Fuzzballs } \\

\vspace{8ex}

{\large  S. Kalyana Rama}

\vspace{3ex}

Institute of Mathematical Sciences, C. I. T. Campus, 

Tharamani, CHENNAI 600 113, India. 

\vspace{1ex}

email: krama@imsc.res.in \\ 

\end{center}

\vspace{6ex}

\centerline{ABSTRACT}
\begin{quote} 

  Assume that there exists a fundamental theory of gravity which
  has a fundamental length scale and which is capable of
  explaining black hole evolution process fully in terms of
  fundamental microscopic degrees of freedom. The resultant
  evolution will clearly be unitary and will have no information
  loss. Carrying this assumption to its logical conclusion and,
  in particular, by considering the size of the region obtained
  by resolving the black hole singularity in such a theory, we
  arrive at a scenario very similar to that in Mathur's fuzz
  ball proposal where there is no horizon. We also comment on
  the consistency of such a picture with the recent results of
  Almheiri et al.

\end{quote}

\vspace{2ex}


\newpage

{\bf 1.}  
We start by assuming that there exists a fundamental theory of
gravity, such as string theory or loop quantum gravity or spin
foams or spin networks, which has a fundamental length scale and
which is capable of explaining black hole evolution process
fully in terms of fundamental microscopic degrees of
freedom. The resultant evolution will clearly be unitary and
will have no information loss.

We carry this assumption to its logical conclusion. Thus, black
hole singularity will be resolved in such a theory.
Furthermore, it is physically reasonable to expect that such a
resolution in terms of microscopic degrees of freedom will
contain information about various constituents that went into
the black hole and, in the semiclassical approximation,
disappeared down its singularity. We then consider the possible
size of the region obtained by resolving the black hole
singularity, and argue that it is parametrically larger than
Planck length.  Analysing it further, we arrive at a scenario
very similar to that in Mathur's fuzz ball proposal where there
is no horizon \cite{fuzz}.

We then comment on the consistency of such a scenario with the
recent results of Almheiri et al \cite{amps}.  Essentially,
statements (ii) and (iii) given in the abstract of their paper
are not true in our scenario. We also point out one challenge
(among many others) that must be faced in our scenario or in
fuzzball proposal. This challenge is about the construction of a
Oppenheimer -- Volkoff type static solution for a star of
arbitrary mass, its stability against collapse, and its entropic
content.

In section {\bf 2}, we present our main arguments and arrive at
a scenario similar to that in fuzz ball proposal. In section
{\bf 3}, we comment on the consistency of such a scenario with
the results of Almheiri et al. We then conclude in section {\bf
4}, by pointing out a challenge for our scenario and for
fuzzball proposal.

\vspace{4ex}

{\bf 2.}  
Consider the formation of a black hole by the collapse of, for
example, a massive star in a pure state. As per the standard
picture, such a black hole has a central singularity, has a
horizon which covers it, emits Hawking radiation from its
horizon, becomes smaller, and evaporates away completely in a
finite time, leaving only the Hawking radiation. Such an
evolution is non unitary since a pure state evolves to a mixed
state, and leads to information loss.

Consider a fundamental theory of gravity, such as string theory
or loop quantum gravity or spin foams or spin networks, which
has a fundamental length scale and which is assumed to be
capable of explaining black hole evolution process fully in
terms of fundamental microscopic degrees of freedom. The
resultant evolution will clearly be unitary and will have no
information loss. Such a theory should, beside other things,
also be able to explain Hawking radiation and resolve black hole
singularity.

\begin{itemize}

\item

Indeed, string theory explains black hole entropy and Hawking
radiation as arising from low energy excitations living on
intersecting branes. Developing these ideas further, Mathur has
proposed a fuzz ball picture for black hole evolution
\cite{fuzz}. Its salient feature, for our purposes here, is that
there is no horizon in the fuzz ball description of black holes.

\item

Ashtekar and Bojowald, and also Hayward, consider the resolution
of black hole singularity by loop quantum gravity, and propose a
paradigm for black hole evolution \cite{ab}. Its salient
feature, again, is that there is no event horizon but only an
apparent horizon.  Thus, collapsing matter and other things that
go into black hole can eventualy escape to outside regions. See
the corresponding Penrose diagrams in these works. Assuming that
one understands the singularity resolution in sufficient detail,
the black hole evolution may be described unitarily. According
to these works, information escapes near the end of evaporation.
If this is true then one encounters `remnant problems', but see
below.

\item

Stephens, 't Hooft, and Whiting have also studied black
evolution assuming that it is unitary, and argue that there will
be no information loss \cite{sthw}. They then describe the
Penrose diagram for the evolution process. The salient feature
here, again, is the absence of an event horizon.

\item

Several years ago, Hawking himself has argued for the unitary
evolution of black hole and for no information loss \cite{h05}.
According to Hawking's quote given in the first paper of Hayward
in \cite{ab}, this seems to come about because `... a true event
horizon never forms, just an apparent horizon.'

\item

Wald has always argued that the presence of an event horizon
will lead to information loss and non unitary evolution
\cite{w}. This is because the causal structure inside the event
horizon, by definition, prevents the inside information from
reaching outside.

\end{itemize} 

The above points all arise in the course of attempting to
describe black hole evolution process in terms of fundamental
microscopic degrees of freedom. They strongly suggest that if an
event horizon forms in the process of black hole formation then
the subsequent evolution of that black hole will be non unitary
and information will be lost. A straightforward corollary then
is that if the black hole formation and evolution process is to
be unitary then no event horizon may form. An apparent horizon
may or may not form, its formation is not a necessary part of
the corollary.

One often assumes that the distinction between event and
apparent horizons becomes important only near the end of
evaporation process when the radius of spacetime curvature and
the black hole size are of order of a few Planck lengths. Until
this end stage, the standard description given by the
semiclassical action is expected to be valid. In particular,
just as for event horizons, nothing from inside the horizon is
expected to reach outside until near the end. It then follows
that information about matter that formed the black hole, and
about the Hawking photons which went inside the horizon during
evaporation, must all escape from an end-stage object of
Planckian size. The amount of such iformation depends on the
initial size of the black hole and may be arbitrarily
large. Hence, its storage in, and escape from, Planckian sized
objects lead to various conundrums which are referred to as
`remnant problems'.

Let there be a theory which resolves, or is capable of
resolving, the black hole singularities. Let $l_f$ be the
fundamental length scale for this theory, far above which one
obtains the standard semiclassical action. For example, $l_f
\sim l_{pl}$, the Planck length, in loop quantum gravity, spin
foam, or spin network theories; whereas $l_f \sim l_s$ in string
theory. If string theory coupling constant $g_s \sim {\cal
O}(1)$, as may perhaps be required for singularity resolution,
then $l_s \sim l_{pl} \;$. Here, we take $g_s \sim {\cal O}(1)$
and set $l_f = l_{pl} \;$.

Let $l_{cld}$ be the size of the region obtained by resolving
the singularity. Then, $l_{cld}$ is the length scale where the
semiclassical action describing the black hole evolution process
should be corrected and where the distinction between event and
apparent horizons becomes important. Also, if the size of black
hole horizon becomes comparable to $l_{cld}$ then information
from inside the horizon may be expected to reach outside. One
may therefore say that there is no horizon when horizon size and
$l_{cld}$ are comparable.

It is generally assumed that $l_{cld} \sim l_{pl} = l_f
\;$. Now, in a spacetime obtained from a semiclassical action,
one expects the action to be corrected when the spacetime
curvature approaches Planckian strengths. Within the
semiclassical calculations, this typically happens a few Planck
lengths away from the singularity. One should then use the
fundamental theory and resolve the singularity. Such a
resolution, given by the fundamental theory in terms of its
microscopic degrees of freedom, must contain information about
what all had gone into the singularity. Logically, it does not
then follow that the size of the region obtained upon resolving
the singularity will also be of order Planck length. Hence, we
denote the size of the resolved region by $l_{cld} \;$. Also,
for the sake of convenience, we refer to the resolved region as
singularity cloud.

This singularity cloud resides at the center of a black hole and
is made up of objects from fundamental theory -- such as stringy
excitations, or Ashtekar loops, or spins of spin foams or spin
networks. Although more things may fall into the cloud, interact
with its constituents, and get transformed to fundamental
objects, nothing from the cloud may escape and reach outside the
black hole horizon until the black hole size and $l_{cld} \;$
are comparable.

We now argue that $l_{cld}$ must be parametrically larger than
the fundamental length scale $l_f$, must depend on the initial
black hole mass $M_{init}$, and must be of the form
\begin{equation}\label{lcld}
l_{cld} \stackrel{>}{_\sim} (l_f \; M_{init})^\alpha \; l_f
\end{equation} 
where it is physically reasonable to expect that $\alpha \ge
\frac{1}{d - 1}$ for a $d-$dimensional black hole. \footnote{
  However, even if $\alpha < \frac{1}{d - 1} \;$, the arguments
  given below and the conclusions they lead to are all valid as
  long as $\alpha$ is strictly positive. The validity of our
  arguments requires only that $l_{cld}$ be dependent on
  $M_{init}$ and be parametrically larger than $l_f$.} Consider
the initial constituents which went in to form the black hole of
mass $M_{init} \;$. Upon black hole formation, these
constituents are taken to have disappeared down the singularity.
But they must show up when the singularity is resolved using the
fundamental theory. After this resolution, the initial
constituents may be thought of as being transformed mostly to $N
\sim l_f M_{init}$ number of fundamental objects each of whose
mass is $\sim l_f^{- 1} \;$. It is then reasonable to expect
that the closest that $N$ number of fundamental objects, each of
whose mass is $\sim l_f^{- 1} \;$, can be packed is in a spatial
region of size $\sim N^{\frac{1}{d - 1}} \; l_f \;$. Considering
that the actual singularity resolution may not correspond to the
closest packing, one obtains the estimate in equation
(\ref{lcld}). This estimate is analogous to saying that the
smallest region where one mole of hydrogen can be packed is
given by $N_a^{\frac{1}{3}} \; \lambda_{pr}$ where $N_a$ is the
Avagadro's number and $\lambda_{pr}$ is proton's Compton
wavelength. In the case of hydrogen, of course, we know that a
more realistic estimate is vastly larger than this, and is
obtained by replacing $\lambda_{pr}$ by Bohr radius.

Once the black hole forms, it starts to emit Hawking radiation.
A pair of photons tunnel out of vacuum near the horizon; one of
them, the $out-$photon, goes outside to infinity; and the other,
the $in-$photon, falls inside onto the singularity. Now, this
$in-$photon is supposed to carry negative energy and is supposed
to reduce the black hole mass. Then, should $M_{init}$ in
equation (\ref{lcld}) be perhaps replaced by $M_{now}$, the net
black hole mass $now$ at the time of observation?

It is not clear how such an `energy annihilation' is to take
place; and, whether and how it would result in the shrinking of
the singularity cloud. By contrast, a `charge annihilation' is
simple: Negative charges fall onto a positively charged cloud.
In a typical process, a negative and a positive charge meet and
cancel each other's charges and emit a photon. Such a photon may
escape to infinity, or may be confined within a finite region.
Note that the negative and positive charges need not necessarily
disappear by annihilating each other. For example, they may form
an `atomic' bound state; or, negative charges may swarm around
positive charges anchored at lattice sites. Thus, although the
net charge is reduced, the charges themselves may not
necessarily disappear.

Going by this analogy, the Hawking $in-$photons will fall onto
the singularity cloud and will interact there with the matter
quanta that initially formed the black hole. The resultant
product, if any and whose nature is not clear to us, must stay
within the singularity cloud. Until the blackhole size and
$l_{cld}$ become comparable, the matter quanta, the
$in-$photons, and the resultant interaction and decay products
must all stay within the cloud. At the face of it, it seems that
all such processes can only increase the size of the cloud. We
may well be biased in thinking so, but it is not clear to us
what processes may decrease the size of the cloud, except the
process whereby black hole and cloud sizes become comparable to
each other and objects from the cloud may escape to outside.
Therefore, we assume that the inflow of Hawking $in-$photons, as
well as anything else that falls into the black hole, will only
increase the size of the cloud or, at best, keep it unchanged.
\footnote{See \cite{mp, bit} for some recent work on bit models
of Hawking evaporation where similar issues arise. }

We are now led to an interesting situation. Imagine that we
observe a black hole now, which was formed earlier. Its mass is
$M_{now}$ now, and was $M_{init}$ at the time of formation. In
the description of this black hole by a fundamental theory,
there are three length scales now: The two expected ones, namely
$l_f = l_{pl}$ and the horizon size $\sim (l_{pl} \;
M_{now})^{\frac{1}{d - 3}} \; l_{pl} \;$, which already appear
in the semiclassical description; and a new third scale
$l_{cld}$ corresponding to the size of the singularity cloud
which appears only after the singularity is resolved in terms of
fundamental microscopic degrees of freedom. Note that $l_{cld}$
depends on $M_{init}$ as given in equation (\ref{lcld}), and
that $M_{init}$ could be arbitrarily larger than $M_{now}$. It
may be that $M_{init} = 10 \; M_{now}$, or it may equally well
be that $M_{init} = 10^{100} \; M_{now} \;$.  So, it seems that
one cannot predict this third scale $l_{cld}$ from the given
data, namely the mass of the black hole $M_{now} \;$.

One can then envision some strange possibilities. For example,
$M_{init}$ is so large that $l_{cld} \gg (l_{pl} \;
M_{now})^{\frac{1}{d - 3}} \; l_{pl} \;$, {\em i.e.} the
singularity cloud is larger than the horizon size now.
Therefore, this black hole must be emitting information about
the initial matter in the form of fundamental objects from the
singularity cloud; hence, effectively, it has no horizon.  Or,
$l_{cld}$ may be one half or one tenth of the present horizon
size. This would imply that the present horizon will be
`covered' by the singularity cloud, and will disappear at a time
which can not be predicted until the actual time of
disappearance.

Note that our estimate of the size $l_{cld}$ of the singularity
cloud is physically reasonable, conservative, and rests on the
existence of a fundamental theory which is capable of explaining
black hole evolution process fully in terms of microscopic
degrees of freedom. It is also reasonable that the singularity
cloud stores the information about the initial matter. It then
seems that the above strange features arise because of the
assumption that nothing can escape from the cloud to outside
until the horizon size becomes comparable to $l_{cld}$.

The simplest way to eliminate such strange features and regain
predictability is to assume that the estimate for $l_{cld}$
given in equation (\ref{lcld}) is a vast underestimate if
$\alpha \le \frac{1}{d - 1} \;$. It can further be seen that
taking $\frac{1}{d - 1} < \alpha < \frac{1}{d - 3}$ will not
suffice, but that taking $\alpha = \frac{1}{d - 3}$ will suffice
to eliminate the strange features and regain predictability.
This is because $\alpha = \frac{1}{d - 3}$ implies that the
scale $l_{cld}$ is same as the initial horizon scale. Hence,
objects from the cloud can escape to outside right from the
moment of black hole formation. \footnote { Avery, Chowdhury,
and Puhm in \cite{bou} arrive at a similar conclusion from a
very different point of view.} This effectively means that there
is no horizon. Also, since objects from the cloud can escape to
outside, it follows that the size of the singularity cloud will
keep decreasing and remain comparable to black hole size
throughout its evolution.

Such a description of black hole evolution is very similar to
Mathur's fuzz ball proposal. Here we have argued that this must
be the case in any fundamental theory which has a fundamental
length scale and which is capable of explaining black hole
evolution process fully in terms of fundamental microscopic
degrees of freedom. The unitarity of evolution and the absence
of information loss will now follow `trivially' just as for any
lump of burning coal.

\vspace{4ex}

{\bf 3.}  
We now comment on the consistency of our scenario with the
results of Almheiri et al \cite{amps}. As expressed concisely in
their abstract, Almheiri et al ``argue that the following three
statements cannot all be true: (i) Hawking radiation is in a
pure state, (ii) the information carried by the radiation is
emitted from the region near the horizon, with low energy
effective field theory valid beyond some microscopic distance
from the horizon, and (iii) the infalling observer encounters
nothing unusual at the horizon.''

Our scenario is similar in spirit to Mathur's fuzz ball
proposal, but is argued to be applicable for any fundamental
theory which can explain black hole evolution in terms of
microscopic degrees of freedom. In our scenario, statements (ii)
and (iii) of Almheiri et al are not true. The information is
stored in, and emitted to outside from, the singularity cloud
whose size is argued to be comparable to black hole size; and,
the dynamics of objects in the cloud are to be described by the
fundamental theory in terms of microscopic degrees of freedom.
Hence, statement (ii) is not true. Also, an infalling observer
will be falling through the singularity cloud and, depending on
the details such as the density of the cloud and the energy and
the nature of the probes used, the infalling observer may see
things unusual while falling through the black hole. Hence,
statement (iii) is not true.

The singularity cloud here may perhaps be thought of as similar
to the `firewall' of Almheiri et al, but only in as much as
infalling objects interact with the cloud and get `burnt' down
to fundamental objects. But these are really two different
things. The `firewall' stays behind or coincides with the
horizon, which is a well defined null surface, and the time of
its formation is still under debate. The singularity cloud is a
lot more similar to fuzz ball. Just like fuzz ball, it is argued
to be of same size as black hole. Furthermore, there strictly is
no horizon since objects from the cloud can escape to outside.
Also, the cloud is argued to have been formed at the time of
black hole formation.

Our scenario will, probably, be deemed to be far from
conservative. But the physical motivations behind it are the
standard ones: requiring the evolution to be unitary,
information to be not lost, and assuming that a fundamental
theory with a fundamental length scale exists which is capable
of explaining black hole evolution process fully in terms of
microscopic degrees of freedom. It is this last motivation that
is not often invoked, nor carried to its logical conclusion.
String theory is one such fundamental theory and, using it,
Mathur indeed has been led to his fuzzball proposal.

Also, note that if an event horizon is assumed to have been
formed then, by definition and as Wald has always argued,
information will be lost and evolution will be non unitary. Of
course, as has been pointed out often, a simple way out is that
an event horizon does not form, only an apparent horizon
forms. This, together with a few standard assumptions about
singular regions, then leads to `remnant problems'. One is then
led to issues related to firewalls, complementarity, et cetera
as can be seen in the works of Almheiri et al and several other
works based on it \cite{bou}. With the physically motivated
present assumptions about singular regions, these issues are all
bypassed.

Use of holographic principle to understand black hole evolution
is an interesting approach, particularly with Maldacena's
conjecture relating gravity in $AdS_5$ to a four dimensional
Yang-Mills (YM) theory. It is known that a black hole in $AdS_5$
corresponds to YM theory at finite temperature \cite{wi}, and
that absorption by black hole corresponds to thermalisation in
YM theory \cite{ks}. However, to the best of our knowledge, it
is not yet known as to {\em e.g.} what an infalling observer
sees; or, whether Mathur's fuzz ball proposal is correct, and
whether horizon is absent; or, how the black hole singularity
gets resolved.

\vspace{4ex}

{\bf 4.}
We conclude by pointing out one challenge (among many others)
that must be faced in our scenario, or in fuzzball proposal,
where it is argued that there is no horizon when a black hole is
described by a fundamental theory in terms of fundamental
microscopic degrees of freedom. The horizon arises only in the
description by a semiclassical theory.

Consider a star. Following Oppenheimer and Volkoff, one can
construct static star solutions assuming an equation of state
for the constituent matter. Typically, the radius of the star
increases with its mass and the mass-radius relation can be made
linear for some parameter values. However, in all such solutions
known so far, an instability sets in for sufficiently high mass
and/or sufficiently high central density. This instability
causes the star to collapse. Hence, it is believed that any
sufficiently massive star will collapse, and that the final
collapse product will be a standard black hole with horizon
\cite{st}.

Our present scenario, or fuzzball proposal, argues that there is
no horizon. Then, using the microscopic degrees of freedom of
the fundamental theory, {\bf (1)} it must be possible to
describe the singularity cloud or fuzz ball by an effective
equation of state; to construct an Oppenheimer-Volkoff type
static star solution; to show that the size of such a star is of
the order of its Schwarzschild radius; and, most importantly, to
show that there is no instability for any value of mass and/or
central density. Moreover, {\bf (2)} the relevant equation of
state for the constituent fundamental objects must be such that
the entropy of such a star is comparable, if not exactly equal,
to the entropy of a corresponding standard black hole.

Obtaining these two results will perhaps make the present
scenario or fuzz ball proposal more compelling. At present, we
do not know how to prove or disprove these results but we
believe that it is a goal worth pursuing.

\vspace{4ex}

{\bf Acknowledgement:} 

We thank G. Date, R. Kaul, P. Mukhopadhyay, and B. Sathiapalan
for discussions. The point mentioned in the first footnote came
up during a seminar in IMSc, Chennai. We thank the seminar
participants for comments and discussions.



\end{document}